\documentclass[preprint, onecolumn,showpacs,preprintnumbers,amsmath,amssymb,floatfix]{revtex4}
\usepackage{graphicx}

\begin{document}

% \draft

\title{Gated nonlinear transport in organic polymer field effect transistors}

\author{B.H. Hamadani and D. Natelson}

\affiliation{Department of Physics and Astronomy, Rice University, 6100 Main St., Houston, TX 77005}

\date{\today}

\begin{abstract}

We measure hole transport in poly(3-hexylthiophene) field effect
transistors with channel lengths from 3~$\mu$m down to 200~nm, from
room temperature down to 10~K.  Near room temperature effective
mobilities inferred from linear regime transconductance are strongly
dependent on temperature, gate voltage, and source-drain voltage.  As
$T$ is reduced below 200~K and at high source-drain bias, we find
transport becomes highly nonlinear and is very strongly modulated by
the gate.  We consider whether this nonlinear transport is contact
limited or a bulk process by examining the length dependence of linear
conduction to extract contact and channel contributions to the
source-drain resistance. The results indicate that these devices are
bulk-limited at room temperature, and remain so as the temperature is
lowered.  The nonlinear conduction is consistent with a model of
Poole-Frenkel-like hopping mechanism in the space-charge limited
current regime.  Further analysis within this model reveals
consistency with a strongly energy dependent density of (localized)
valence band states, and a crossover from thermally activated to
nonthermal hopping below 30~K.

\end{abstract}

\maketitle 

\newpage

\section*{I. INTRODUCTION}

Electronic devices based on organic semiconductors, such as polymeric
field effect transistors (FETs) and light emitting diodes (LEDs), have
attracted much interest as possible inexpensive and flexible
alternatives to inorganic
devices\cite{DimitrakopoulosetAl01IBM,BaoetAl99JMC}.  While there has
been considerable improvement in device properties, the detailed
mechanism of electronic transport in organic thin film devices remains
a subject of active investigation.

Charge motion in undoped organic field effect devices is often
characterized in the linear (small source-drain bias) regime by a
field-effect mobility.  In disordered polymer semiconductors this
mobility, $\mu$, is strongly temperature dependent near room
temperature, consistent with thermally assisted hopping between
localized states dispersed throughout the polymer
film\cite{Hoppingbook}.  These localized states are likely to be
polaronic\cite{ParrisetAl01PRL}.  Several different models for hopping
transport in these materials have been used to interpret experimental
data from $\sim$200~K to room temperature, including simple thermal
activation\cite{BrownetAl97SM}, 2d variable range hopping
(VRH)\cite{AleshinetAl01SM}, and percolative VRH with an exponential
density of states\cite{VissenbergetAl98PRB}.  While some models work
better with some specific samples, generally distinguishing between
them is difficult.

The situation is complicated by the fact that the effective mobilities
inferred in this manner depend on $T$, gated charge
density\cite{DimitrakopoulosetAl99Science}, source-drain bias, and
contact effects.  Parasitic contact resistances in particular can be
important.  Experiments in several organic semiconductor/electrode
combinations\cite{HorowitzetAl99JAP,SeshadrietAl01APL,StreetetAl02APL,BurgietAl02APL,KaluketAl03SSE,NecludiovetAl03SSE,ZaumsiletAl03JAP,MeijeretAl03APL}
have shown that contact resistances can be a significant fraction of
the total source-drain resistance in the linear regime in short
devices.  Data in this shallow channel limit must be corrected
accordingly for this parasitic series resistance, $R_{\rm s}$, to find
the true mobility within the semiconductor.  As with $\mu$, one must
bear in mind that $R_{\rm s}$ typically depends on temperature, gate
voltage, and local electric field\cite{BurgietAl02APL}.  When
examining transport properties of organic FETs, it is therefore
important to determine whether the devices are dominated by the bulk
(channel) or the contacts.

We report {\it nonlinear} transport measurements in field-effect
devices made from high quality, solution cast, regio-regular
poly(3-hexylthiophene) (P3HT), with channel lengths from 3~$\mu$m to
200~nm and aspect ratios of 10.  Higher temperature properties at low
source-drain fields are consistent with those observed by other
investigators.  From 200~K to 10~K, we observe gate-modulated
nonlinear $IV$ characteristics.  To understand the effects of contacts
in our series of devices, we examined both the devices described
above, and an additional series of fixed-width FETs to obtain the
channel and contact resistances as a function of temperature.  These
data demonstrate that (a) the fixed-apect-ratio devices are bulk (not
contact) limited at high temperatures; and (b) the contribution of
contacts relative to the channel actually {\it decreases} as the
temperature is lowered, so that bulk-limited devices tend to remain so
as $T$ is decreased.  Since the nonlinearities in the $IV$ curves
become more pronounced at low temperatures and in wider devices (for
which contact resistance are proportionately less important), it is
unlikely that these nonlinearities are due to contact effects in these
geometries.  We find that the nonlinear data are consistent with a
model of Poole-Frenkel (PF) type conduction in the space-charge
limited (SCL) regime.  Within this framework, the strong gate and
temperature dependence of this conduction are consistent with a
density of localized valence states that varies approximately
exponentially in energy.  Still within this model, at temperatures
below 30~K there appears to be a crossover from thermally assisted
hopping to a nonthermal mechanism.  These nonlinear data over a broad
temperature range constrain any other models of transport in such
devices.  Studying FETs in this nonlinear regime allows comparisons
between models not readily performed with linear transport data.

\section*{II. EXPERIMENTAL DETAILS}

Devices are made in a bottom-contact configuration (see
Fig.~\ref{fig:mobT} inset) on a degenerately doped $p+$ silicon wafer
to be used as a gate.  The gate dielectric is 200~nm of thermal
SiO$_{2}$.  Source and drain electrodes are patterned using standard
electron beam lithography.  The electrodes are deposited by electron
beam evaporation of 4~nm Ti and 25~nm of Au followed by liftoff.  The
fixed-aspect-ratio devices have channel lengths ranging from 3~$\mu$m
down to 200~nm, with the channel width scaled to maintain $w/L = 10.$
Larger FET devices ($w =$~1~mm, $L =$~50~$\mu$m) are also prepared for
comparison.  We also examine a second set of devices with fixed width
$w=100~\mu$m and channel lengths varying from 5~$\mu$m up to
40~$\mu$m, specifically for probing contact resistance issues.

The organic semiconductor is 98\% regio-regular P3HT\cite{purchase}, a
well studied
material\cite{BaoetAl96APL,SirringhausetAl98Science,SirringhausetAl99Nature}.
P3HT is known from x-ray scattering to form nanocrystalline domains
with sizes on the order of
20~nm\cite{BaoetAl96APL,SirringhausetAl99Nature}, and more ordered
films correlate with higher measured mobilities.  RR-P3HT is dissolved
in chloroform at a 0.02\% weight concentration, and is solution
cast\cite{BaoetAl96APL} onto ozone-cleaned, choloroform-swabbed
substrates.  The fixed-aspect-ratio series of devices are from one
casting, while the fixed-width series are from a second casting.  The
resulting film thicknesses over the channel region are tens of nm as
determined by atomic force microscopy (AFM), though casting produces
somewhat nonuniform films.  All devices are stored in vacuum
dessicators until use.  The measurements are performed in vacuum
($\sim 10^{-6}$~Torr) in a variable-temperature probe station using a
semiconductor parameter analyzer (HP4145B).

\section*{III. RESULTS AND DISCUSSION}

At room temperature, the devices operate as standard $p$-type FETs in
accumulation
mode\cite{BaoetAl96APL,SirringhausetAl98Science,SirringhausetAl99Nature,AleshinetAl01SM,MeijeretAl02APL}. With
the source electrode as ground, in the linear regime we
extract\cite{BrownetAl97SM} an effective mobility from the
transconductance.  That is, from data of source-drain current, $I_{D}$
versus the gate voltage, $V_{G}$, at a fixed low drain voltage,
$V_{D}$, we compute $\mu=(g_{m}L)/(wC_{i}V_{D})$, where $g_{m} \equiv
\partial I_{D}/\partial V_{G}$ is the transconductance, $C_{i}$ is
capacitance per unit area of the gate insulator, and $w/L$ is the
aspect ratio.  The relevance of parasitic contact resistances will be
addressed below.  As is reported elsewhere\cite{HorowitzetAl99JAP},
the mobility is gate voltage dependent, increasing with increasing
$V_{G}$.  It also increases with increasing source-drain voltage.
Effective mobilities are typically between $10^{-3}$ and $10^{-2}$
cm$^{2}$/Vs, and apparent threshold voltages ($V_{T}$), though not
necessarily meaningful\cite{MeijeretAl02APL}, are low ($<2$ V).  For
$L=$~50~$\mu$m FETs operated in the saturation regime, the on/off
ratio is typically $\sim~650$, comparing between gate voltages of
-95~V and 0~V.  As temperature is reduced below 300~K, the off-current
drops to undetectable levels by 150~K, as the unintentional carriers
(due to slight doping from air exposure) freeze out.

Over the moderate temperature range of 300~K to $\sim$ 200~K, the
mobility as inferred above at fixed small $V_{D}$ is found to depend
steeply on temperature.  A small representative set of this data is
shown in Fig.~\ref{fig:mobT}, where $\mu$ is plotted vs. inverse
temperature for $V_{G}= -30, -40,$~ and -50~V, for the $L=300$~nm,
$w=3~\mu$m device at constant source-drain electric field of $1.3
\times 10^{7}$ V/m.

These data are approximately equally consistent with the three models
mentioned above: simple Arrhenius behavior (the dashed line) with an
activation energy $\sim$ 100~meV; VRH for a 2-D system (dotted
line), of the form $\mu=\mu_{0}\exp(-(T_{0}/T)^{1/3})$, where
$\mu_{0}$ and $T_{0}$ are fit parameters; and finally the more
sophisticated percolative VRH theory (solid lines) developed by
Vissenberg {\it et al}\cite{VissenbergetAl98PRB,MeijeretAl02APL}.  The
Vissenberg model's underlying assumptions include an exponential
density of (localized) states (DOLS),
$\nu(\epsilon)\sim\exp(\epsilon/k_{\rm B}T_{0})$, with transport of
carriers dictated by percolative hopping. Since the gate voltage
controls the Fermi level in the channel, and hence the occupation of
the localized states, one finds that transport in the channel is
strongly affected by $V_{G}$.  Relevant fit parameters
\cite{VissenbergetAl98PRB,MeijeretAl02APL} are: $T_{0}$, describing
the energy dependence of the DOLS; a prefactor $\sigma_{0}$ with units
of conductivity; and $\alpha$, an effective overlap parameter for
hopping.  Values used in the fits shown are $\sigma_{0}=7 \times
10^{5}$~S, $T_{0}=$~418~K, $\alpha = 4.35 \times 10^{9}$~m$^{-1}$,
consistent with those seen by other investigators in
P3HT\cite{MeijeretAl02APL}.

As temperatures are lowered from 200~K down to 10~K, over a broad
range of source-drain and gate voltages, $I_{D}$ evolves from
approximately linear to a strongly nonlinear (superquadratic)
dependence on $V_{D}$.  An example of this evolution is shown in
Fig.~\ref{fig:nonlinIV} for the $L = 3~\mu$m, $w = 30~\mu$m device,
comparing data at 300~K and 70~K.  We note that, at the lowest
temperatures, smaller devices transport current {\it more easily}
(larger currents at smaller gate voltages for a fixed $V_{D}/L$) than
larger devices, as we will discuss later. 

Analysis below shows that the nonlinear $IV$ characteristics are
decribed well by a model incorporating space-charge limited currents,
modified by a Poole-Frenkel-like exponential dependence of effective
mobility on square root of the local electric field (SCLPF).  This
conduction mechanism has been seen repeatedly in {\it two}-terminal
devices\cite{BlometAl97PRB,MalliarasetAl98PRB}.  Room temperature
experiments\cite{AustinetAl02APL} on P3HT FETs with 70~nm channel
lengths also show indications of SCL currents.  We find that within
this model, the temperature and gate voltage dependence of the data
support a strongly energy dependent DOLS such as that in the
Vissenberg picture.  Other models may be possible, but they are
constrained by the dependences presented below.

Charge transport in a device is space charge limited if the injected
carriers significantly alter the local electric field from the average
field imposed by the electrode potentials, and correspondingly limit
the current.  If, instead, the bottleneck in charge transport is
injection at the contacts, a device is said to be contact limited, and
is expected to exhibit Ohmic behavior at low source-drain fields.  In
principle, modeling our devices requires the full solution of the
steady state charge and electric field profile in a three-terminal
accumulation mode transistor, including field- and temperature
dependent effective mobility, and field- and temperature dependent
contact properties.  This general problem is very
complex\cite{AlametAl97IEEE}; here we consider a simpler model and
compare with the transport data.

In a system that is not contact limited, when the effective mobility
varies as a function of electric field $F$, the space-charge limited
current in a {\it two-terminal} device (a 1d model) is determined by the
numerical solution of
\begin{eqnarray}
J &=& p(x)e \mu(F(x))F(x), \nonumber\\
\frac{\kappa \epsilon_{0}}{e}\frac{dF}{dx}& = &p(x),
\label{eq:coupled}
\end{eqnarray}
where $p(x)$ is the local hole density, $\int_{0}^{L}F(x)dx = V$, and
the appropriate boundary condition on $F(x=0)$.  For approximately
Ohmic contacts, $F(x=0)\approx 0$.  

We note that the case of an effective mobility that varies as
\begin{equation}
\mu(F) = \mu_{0} \exp(\gamma \sqrt{F})
\label{eq:PFlike}
\end{equation}
and Ohmic contacts has been
solved\cite{Murgatroyd70JPD}, and that the exact numerical
solution is very well approximated by:
\begin{equation}
I \approx \frac{9}{8}\kappa \epsilon_{0} \mu_{0} \left(\frac{V}{L}\right)^{2}\frac{A}{L}\exp\left[0.9\gamma \left(\frac{V}{L}\right)^{1/2}\right].
\label{eq:SCLPF}
\end{equation}
Here $A$ is the device cross-sectional area, and $L$ is the
interelectrode distance.  In our geometry $L$ is the channel length,
$A$ is an effective cross-sectional area for the device (proportional
to channel width), and $\kappa$ is the relative dielectic constant of
the semiconductor (chosen to be 3 in our analysis).  The appearance of
$V/L$ in this equation does not imply that the electric field is
constant over the device length.  Rather, Eq.~\ref{eq:SCLPF} suggests
a means of plotting $IV$ data to quickly ascertain consistency with
the detailed numerical solution to the two-terminal SCLPF problem.

Note that $\mu_{0}$ can depend on temperature; in a picture of hopping
it should be proportional to the effective DOLS at the injecting
contact.  The dependence of mobility on $\exp(\sqrt{F})$ has long been
seen in semiconducting and conducting polymers\cite{Gill72JAP}, and is
associated with the charge carriers and disorder in these
materials\cite{NovikovetAl95JPC,BlometAl97PRB,ParrisetAl01PRL,RakhmanovaetAl00APL}.
The numerical solution of Ref.~\cite{Murgatroyd70JPD} (approximated by
Eq.~\ref{eq:SCLPF}) should be valid as long as the functional form of
the field-dependent mobility remains $\exp(\sqrt{F})$.

This equation is derived\cite{Murgatroyd70JPD,MalliarasetAl98PRB}
assuming that the charge distribution is determined by source-drain
electrostatics only.  The electric field from the gate certainly plays
a nontrivial role in our devices, clearly affecting charge injection,
and allowing the formation of a channel at higher temperatures (since
any charge present in the channel below dopant freeze-out has to be
injected from the source and drain).  For simplicity, however, in this
model we will assume that the gate dependence will manifest itself
through $\mu_{0}$, and that for fixed gate voltage we may treat the
source-drain conductance like a two-terminal device.

Figure~\ref{fig:SCLPF}a shows a representative log-log plot of $I_{D}$
vs. $V_{D}$ for one sample with $L= 500$~nm and $V_{G}=-75$~V, for
several temperatures.  The solid lines are the numerical solution to
Eq.~\ref{eq:coupled}, with parameters $A\mu_{0}$ and $\gamma$ chosen
at each temperature to give the best fit.  The numerical solution is
virtually indistinguishable from the analytic form of
Eq.~\ref{eq:SCLPF}.  Data on this and other samples for different gate
voltages are qualitatively similar, with very good agreement between
the numerical solution of the SCLPF model and the data.
Fig.~\ref{fig:SCLPF}b shows the same data and fits replotted as
suggested by Eq.~(\ref{eq:SCLPF}).  The quality of this agreement
between the SCLPF model and the data in several devices over a broad
range of $T$, $V_{G}$, and $V_{D}/L$ is striking.  Clearly for a given
(large) $V_{G}$ and (low) $T$, $I_{D}\sim V_{D}^{2}\exp(\sqrt{V_{D}})$.
Alternate explanations of the nonlinear conduction are strongly
constrained by this dependence.  The temperature and gate voltage
range over which this form of source-drain nonlinearity occurs varies
systematically with sample size, as described below.

\subsection*{A. Contact effects} 

One must consider whether the nonlinear $I_{D}-V_{D}$ characteristics
result from nonlinear contact resistances as the temperature is
decreased.  For several reasons, discussed below, we do not believe
this to be the case.

First we consider directly inferring the contact and channel
resistances ($R_{\rm s}$ and $R_{\rm ch}$, respectively) in the linear
regime, and examining the temperature variation of their relative
contributions.  For a series of devices with fixed width, these
resistances are calculated as
follows\cite{KaluketAl03SSE,NecludiovetAl03SSE,ZaumsiletAl03JAP,MeijeretAl03APL}:
At a given $T$, the total resistance, $R_{\rm on}\equiv \partial
I_{D}/\partial V_{D}$, is calculated for each device at a small
$V_{D}$ and is plotted as a function of $L$ for each gate voltage.
The channel resistance per unit length, $R_{\rm ch}/L$, at a given $T$
and $V_{G}$ is the slope ($\partial R_{\rm on}/\partial L$) of such a
graph, and the intercept ($R_{\rm on}$ extrapolated to $L=0$) gives
the parasitic series resistance, $R_{\rm s}$.  

For a fixed-aspect-ratio series of devices, one may follow an
analogous procedure.  The total source-drain resistance $R_{\rm on} =
(L/w)R_{\Box}+R_{\rm s}$, where $R_{\Box}$ is the resistance per
square of the channel.  At a given $T$ and $V_{G}$, $R_{\rm on}\times
w$ is plotted versus $L$ for the series of devices.  The slope of such
a graph gives $R_{\Box}$, and the intercept gives the total parasitic
contact resistivity, $R_{\rm s}\times w$.  This analysis is shown in
the fixed-aspect-ratio devices in Fig.~\ref{fig:olddataRvsL} at room
temperature for several gate voltages.  For our geometry of $w/L=10$,
the inset shows the inferred $R_{\rm s}/R_{\rm ch}$ as a function of
$V_{G}$, for a device with $L=1~\mu$m.  {\it Our series of
fixed-aspect-ratio devices is clearly channel-limited at room
temperature.}

This analysis may be repeated at different temperatures to examine the
evolution of $R_{\rm s}$ and $R_{\rm ch}$.  We find that the nonlinear
conduction at large average source-drain fields shown in the previous
section makes this difficult to measure over a broad temperature range
in the fixed-aspect-ratio device series.  However, the fixed-width
devices with longer channel lengths are well-suited to this approach
down to 100~K.  Figure~\ref{fig:RsRchvsT} shows $R_{\rm s}/R_{\rm ch}$
as a function of temperature for the $L= 5~\mu$m device from the
$w=100~\mu$m series.  Near room temperature, $R_{\rm s}>R_{\rm ch}$
for this device.  As $T$ decreases, while both $R_{\rm s}$ and $R_{\rm
ch}$ increase significantly, $R_{\rm s}$ falls below $R_{\rm ch}$ near
100~K.  The results of this experiment and others to be published
elsewhere\cite{HamadanietAl03sub} demonstrate that the contact
contributions become {\it less} important at low temperatures.  This
strongly suggests that the nonlinearities in the $I_{D}-V_{D}$ curves
of short channel length devices observed at lower temperatures are
unlikely to be contact effects. 

Furthermore, the trends in transport with sample dimensions also
support this conclusion.  We find at low temperatures that the
smallest devices actually transport charge considerably {\it better}
than larger devices.  For example, at 50~K for fixed $V_{D}/L$, the
fixed-aspect-ratio $L=300$~nm device exhibits measurable conduction
for gate voltages as small as -25~V, while the $L=3~\mu$m device
requires $V_{G}=-45$~V.  This trend is the {\it opposite} of what one
would expect for contact-limited conduction\cite{StreetetAl02APL}.
Since $w/L$ is held constant in this set of devices, shorter channel
devices have significantly smaller contact areas as well, further
emphasizing this point.  The data are, however, consistent with the
suggestion of space charge effects seen in 70~nm channel length P3HT
FETs\cite{AustinetAl02APL}.  Finally, the voltage and temperature
dependence of the data in Fig.~\ref{fig:SCLPF} is not consistent with
the forms for either classical Schottky contacts or Fowler-Nordheim
emission.  Detailed theory\cite{ArkhipovetAl98JAP} and
experiments\cite{WoudenberghetAl01APL} on injection into disordered
polymer semiconductors show that injection efficiency can actually
{\it improve} as temperature is decreased, consistent with our contact
resistance data described above. Coupled with the size dependence,
this supports the idea that low temperature transport in our devices
is {\it bulk} limited rather than contact limited.

\subsection*{B. Physical significance of fit parameters}

Continuing within the SCLPF model and our analysis of the
fixed-aspect-ratio devices, 
we note that, for identical effective mobilities and $\gamma$
parameters, Eq.~\ref{eq:SCLPF} implies that two samples with the same
aspect ratio, thickness, and average source-drain field should give
the same currents, independent of channel length, even deep in the
nonlinear regime.  Device-to-device variability in the effective
mobility and $\gamma$, presumably due to differences in P3HT thickness
and microstructure, make this challenging to check directly in our
devices.  For reasonable values of $\gamma$ and $V_{D}/L$, a 10\%
variation in $\gamma$ would lead to more than a factor of 2 variation
in predicted current at low temperatures because of the exponential
dependence in Eq.~\ref{eq:SCLPF}.  However, if one fixes $V_{D}/L$ and
$V_{G}$, and normalizes measured currents by room temperature
mobilities, one does indeed find scaling.  For example, the $L=500$~nm
and $L=1~\mu$m currents in the nonlinear regime normalized this way
agree well all the way down to 10~K.

Within this model, the parameter $\gamma$ should depend only on the
hopping mechanism ({\it e.g.}  thermal activation) and the nature of
the localized states.  We therefore expect $\gamma$ to be independent
of gate voltage for a given sample, and this is indeed seen in the
inset to Fig.~\ref{fig:slopeT}.  At high temperatures ($T > 50$~K),
the data for all gate voltages and all samples look roughly linear in
$1/T$, consistent with thermally activated hopping.  The magnitude of
the slope of $\gamma$ vs. $1/T$ is approximately 0.12~K(m/V)$^{1/2}$.
This is consistent in magnitude with coefficients found in other
semiconducting polymers such as poly(phenylene
vinylene)\cite{BlometAl97PRB}.  However, $\gamma$ vs. $1/T$ deviates
significantly from a straight line at lower temperatures for all
samples.  This is consistent with a crossover from thermally activated
hopping to a much less steep temperature dependence.  A natural
candidate is field enhanced tunneling between the localized states.

We now consider the gate and temperature dependence of the parameter
$A\mu_{0}$ found by the numerical analysis above.  The effective
cross-section for current flow, $A$, is assumed to be temperature and
gate voltage independent for each sample.  In the linear regime at
moderate temperatures it is known that the mobility inferred from the
transconductance is gate voltage dependent, as seen in
Fig.~\ref{fig:mobT}.  This dependence on $V_{G}$ continues in the
apparent SCLPF regime, as shown in Fig.~\ref{fig:intVGT}a on an $L =
300$~nm device.  The amount of variation of $\mu_{0}$ with $V_{G}$,
roughly a 5\% increase of $\mu_{0}$ per volt of $V_{G}$ for this
sample, shows no strong trend with temperature.  The magnitude of this
variation of mobility with $V_{G}$ is consistent with that seen at
higher temperatures in Fig.~\ref{fig:mobT} for this sample.  This
exponential dependence of $\mu_{0}$ on $V_{G}$ is seen throughout the
apparent SCLPF regime.  As in the linear regime of an accumulation
FET, gate voltage modulation of the conduction is {\it unipolar}:
higher currents result only when $V_{G}$ is made more negative.

The temperature dependence of $\mu_{0}$ is also very strong, as shown
in Fig.~\ref{fig:intVGT}b for the same 300~nm channel sample at five
gate voltages.  The variation of $\mu_{0}(T)$ shown is much closer to
an exponential in $T$ to some slightly sublinear power rather than an
Arrhenius or VRH form.  This strong temperature dependence of
$\mu_{0}$ holds for all samples in the apparent SCLPF regime.

The strongly energy dependent DOLS employed in, for example, the
Vissenberg model offers a natural explanation for these steep
dependences of $\mu_{0}$ on $V_{G}$ and $T$.  In the absence of any
gate effect, the {\it effective} density of localized states available
for hopping transport at some temperature $T$ is given by
$\nu(\epsilon \approx k_{\rm B}T)$ where energy is measured from the
band edge.  An exponential DOLS of the Vissenberg model would then
lead to an exponential dependence of $\mu_{0}$ on $T$.  Gate voltage
dependence in this case comes from electrostatic modulation of the
Fermi level in this rapidly varying DOLS.  It is difficult to
understand otherwise how an exponential dependence of the prefactor on
$T$ or $V_{G}$ could arise.  Deriving a quantitative relationship
between $V_{G}$ and the local Fermi level would require solving the
full electrostatic problem of SCPLF conduction in the presence of the
transverse gate field.

One can consider whether the proposed SCLPF conduction is a {\it bulk}
process or one dominated by conduction in the thin channel layer
active in standard FET operation.  For the SCLPF mobility parameter
$\mu_{0}$ to coincide with the zero field mobility found at high
temperatures in the linear regime, an effective cross-sectional area
$A$ considerably larger than $w \times$ a few nanometers is required.
This is also true for the data of Ref.~\cite{AustinetAl02APL}, in
which a considerably different device geometry was used, if analyzed
using the SCLPF model.  However, the model of
Eqs.~(\ref{eq:coupled}),(\ref{eq:PFlike}) does not account for the
presence of a gate electrode or complicated source and drain
geometries, and full computational modeling in this regime may be
required to quantitatively account for this.  One test for bulk
vs. channel conduction would be to search for a correlation between
P3HT thickness and currents in the apparent SCL regime.  This
investigation is ongoing.

\section*{V. CONCLUSIONS}

For a series of field effect devices with channel lengths ranging from
3~$\mu$m to 200~nm, we find gate modulated nonlinear conduction at low
temperatures and high average source-drain electric fields.  Analysis
of channel and contact resistances as a function of temperature, and
the dependence of conduction on sample size at low temperatures
support the conclusion that this nonlinearity is unlikely to be a
contact effect.  We find that the data are well described by a model
of gate modulated space-charge limited currents with
Poole-Frenkel-like behavior of mobility.  Within this model, the
$V_{G}$ and $T$ dependence of the mobility prefactor is consistent
with a very strongly energy dependent density of localized states.
Finally, the temperature dependence of the Poole-Frenkel-like term
within this model suggests a crossover from thermal hopping to quantum
tunneling at low temperatures.  Further studies of the field effect
electrostatics problem, the metal-semiconductor contacts, and the low
temperature nonthermal hopping process should lead to increased
understanding of the conduction processes at work in these materials.

\section*{VI. ACKNOWLEDGMENT}

The authors gratefully acknowledge the support of the Robert A. Welch
Foundation.

\clearpage

\begin{figure}[h!]
%\begin{center}
%\includegraphics[clip, width=7.5cm]{figures/mobT.eps}
%\end{center}
%\vspace{-3mm}
\caption{\small Mobility vs. $T$ as computed from transconductance for three gate voltages, for a device with $L=300$~nm from room temperature down to 200~K.
Lines are fits to various models of hopping transport described in the text.
Inset:  cross-section of device showing bottom contact configuration and definition of channel length, $L$.}
\label{fig:mobT}
\end{figure}

\begin{figure}[h!]
%\begin{center}
%\includegraphics[clip, width=7.5cm]{figures/nonlinIV.eps}
%\end{center}
%\vspace{-3mm}
\caption{\small $I_{D}$ vs. $V_{D}$ for the $L=3~\mu$m, $w=30~\mu$m
device, at 300~K (nearly linear, top) and 70~K (highly nonlinear, bottom).
Curves from the top down correspond to $V_{G}$ values from -95~V to -30~V in
intervals of 5~V.}
\label{fig:nonlinIV}
\end{figure}

\begin{figure}[h!]
%\begin{center}
%\includegraphics[clip, width=7.5cm]{figures/SCLPF.eps}
%\end{center}
%\vspace{-3mm}
\caption{\small (a) Log-Log plot of $I_{D}$ vs. $V_{D}$ for a device
with $L = 500$~nm at $V_{G}=$-75~V.  Solid lines indicate a numerical
solution assuming space-charge limited conduction with a
Poole-Frenkel-like field dependence of the mobility. (b) Plot of
$\ln(I_{D}/V_{D}^{2}$ vs. $\sqrt{V_{D}}$, as suggested by
Eq.~\ref{eq:SCLPF}.  Solid lines are fits to a linear dependence on
$\sqrt{V_{D}}$.  }
\label{fig:SCLPF}
\end{figure}

\begin{figure}[h!]
%\begin{center}
%\includegraphics[clip, width=7.5cm]{figures/olddataRvsL.eps}
%\end{center}
%\vspace{-3mm}
\caption{\small Plot of $R_{\rm on}\times w$ vs. $L$ for the 
fixed-aspect-ratio device series in the linear regime at 300~K
for several gate voltages.  Slopes of the linear fits correspond
to $R_{\Box}$ of the channel, while intercepts correspond to the
parasitic contact resistivity, $R_{\rm s}w$.  Inset: the ratio
$R_{\rm s}/R_{\rm ch}$ for the $L=1~\mu$m, $w=10~\mu$m device.
Clearly this device is {\it not} contact limited at room temperature.}
\label{fig:olddataRvsL}
\end{figure}

\begin{figure}[h!]
%\begin{center}
%\includegraphics[clip, width=7.5cm]{figures/RsRchvsT.eps}
%\end{center}
%\vspace{-3mm}
\caption{\small $R_{\rm s}/R_{\rm ch}$ as a function of temperature as determined for the $w=100~\mu$m, $L=5~\mu$m member of the fixed-width set of devices.  Since this ratio decreases as $T$ is lowered, contacts actually {\it improve}
relative to the channel at low temperatures.  This interesting result will
be discussed more fully elsewhere\protect{\cite{HamadanietAl03sub}}.}
\label{fig:RsRchvsT}
\end{figure}

\begin{figure}[h!]
%\begin{center}
%\includegraphics[clip, width=7.5cm]{figures/slopeT.eps}
%\end{center}
%\vspace{-3mm}
\caption
{\small Inset: Plot of the parameter $\gamma$ vs. $V_{G}$ for the
500~nm sample of Fig.~\ref{fig:SCLPF} at various temperatures
((top-to-bottom) 10~K, 30~K, 50~K, 70~K, 90~K, 120~K, 150~K, 180~K,
210~K) showing that $\gamma$ is roughly gate voltage independent.
Main figure: Plot of $\gamma$ vs. $1/T$ for several samples, with
$\gamma$ averaged over gate voltages for each sample.  Error bars are
standard deviation.  At high temperatures $\gamma$ is expected to vary
linearly in $1/T$, and all samples show a similar slope, $\gamma\times
T \approx 0.12$~(m/V)$^{1/2}$.  Within this model, saturation of $\gamma$ at low temperatures would indicate a crossover from thermal to nonthermal hopping
transport at low temperatures.  }
\label{fig:slopeT}
\end{figure}

\begin{figure}
%\begin{center}
%\includegraphics[clip, width=8cm]{figures/intVGT.eps}
%\end{center}
%\vspace{-3mm}
\caption
{\small Assuming a fixed effective area $A = 3 \times
10^{-13}$~m$^{2}$, (a) Plot of the parameter $\mu_{0}$ from plots like
Fig.~\ref{fig:SCLPF} vs. $V_{G}$ for the 300~nm sample, from 210~K
down to 10~K.  (b) Plot of $\mu_{0}$ vs. $T$ for the same sample, for
5 different gate voltages.  Note that $\mu_{0}$ depends nearly
exponentially on both temperature and gate voltage.}
\label{fig:intVGT}
\end{figure}

\clearpage

\setcounter{figure}{0}

\begin{figure}
\begin{center}
\includegraphics[width=7cm, clip]{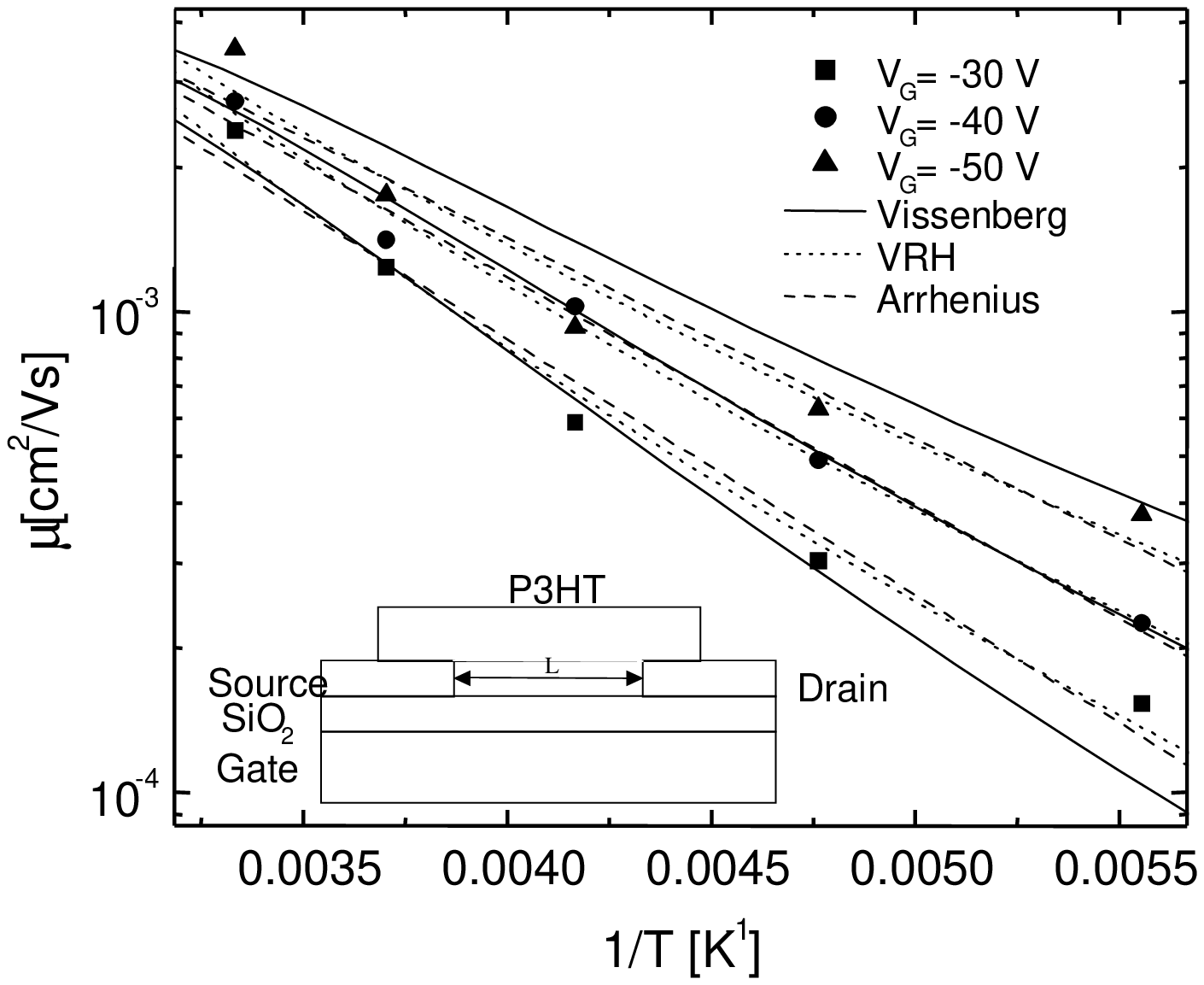}
\end{center}
\vspace{1.5cm}
\caption{of 7, B.H. Hamadani and D. Natelson ``Gated nonlinear transport in organic polymer field effect transistors'', to appear in {\it J. Appl. Phys.}} 
\end{figure} 

\clearpage

\begin{figure}
\begin{center}
\includegraphics[width=7cm, clip]{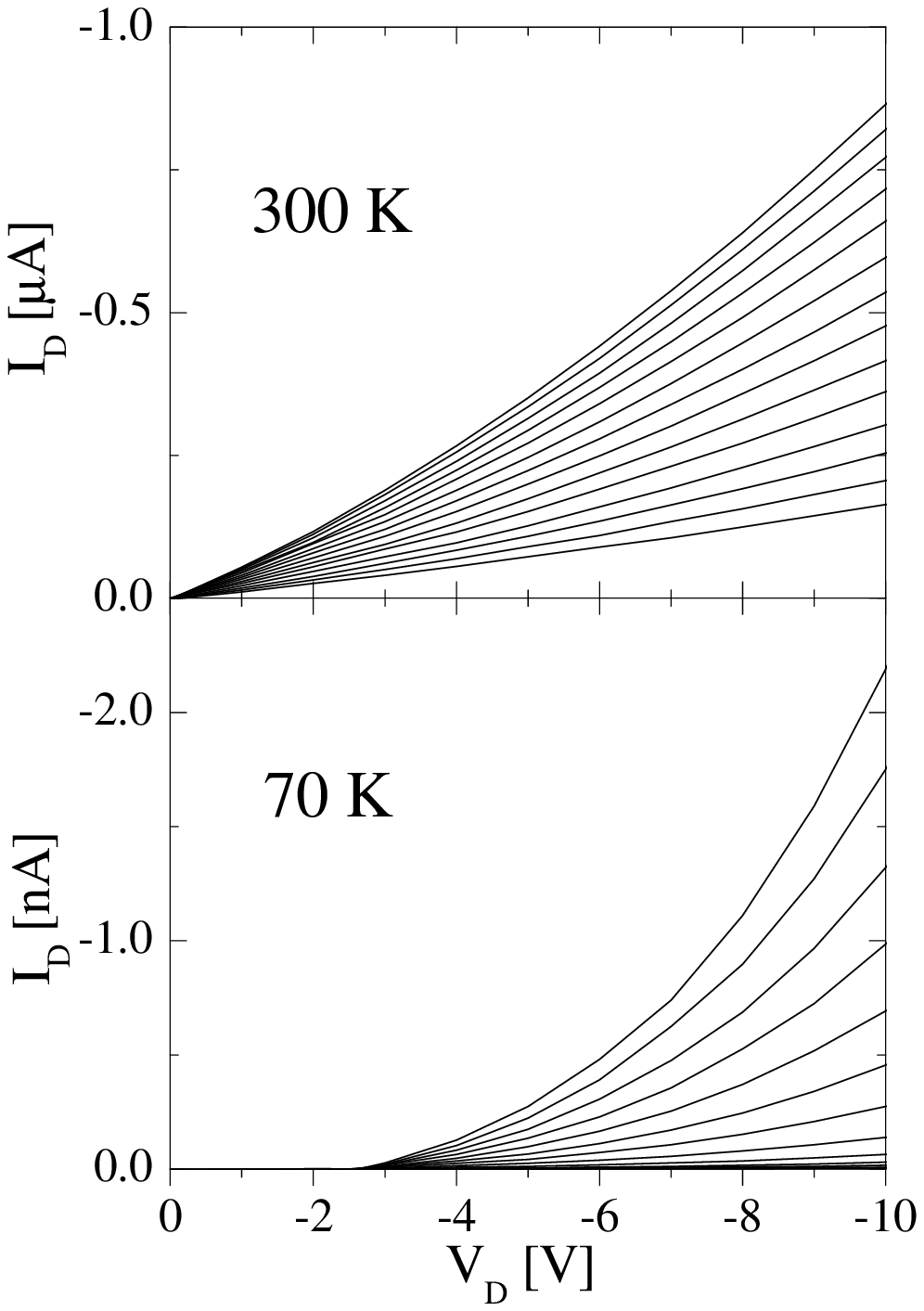}
\end{center}
\vspace{1.5cm}
\caption{of 7, B.H. Hamadani and D. Natelson ``Gated nonlinear transport in organic polymer field effect transistors'', to appear in {\it J. Appl. Phys.}} 
\end{figure} 

\clearpage

\begin{figure} 
\begin{center}
\includegraphics[width=7cm, clip]{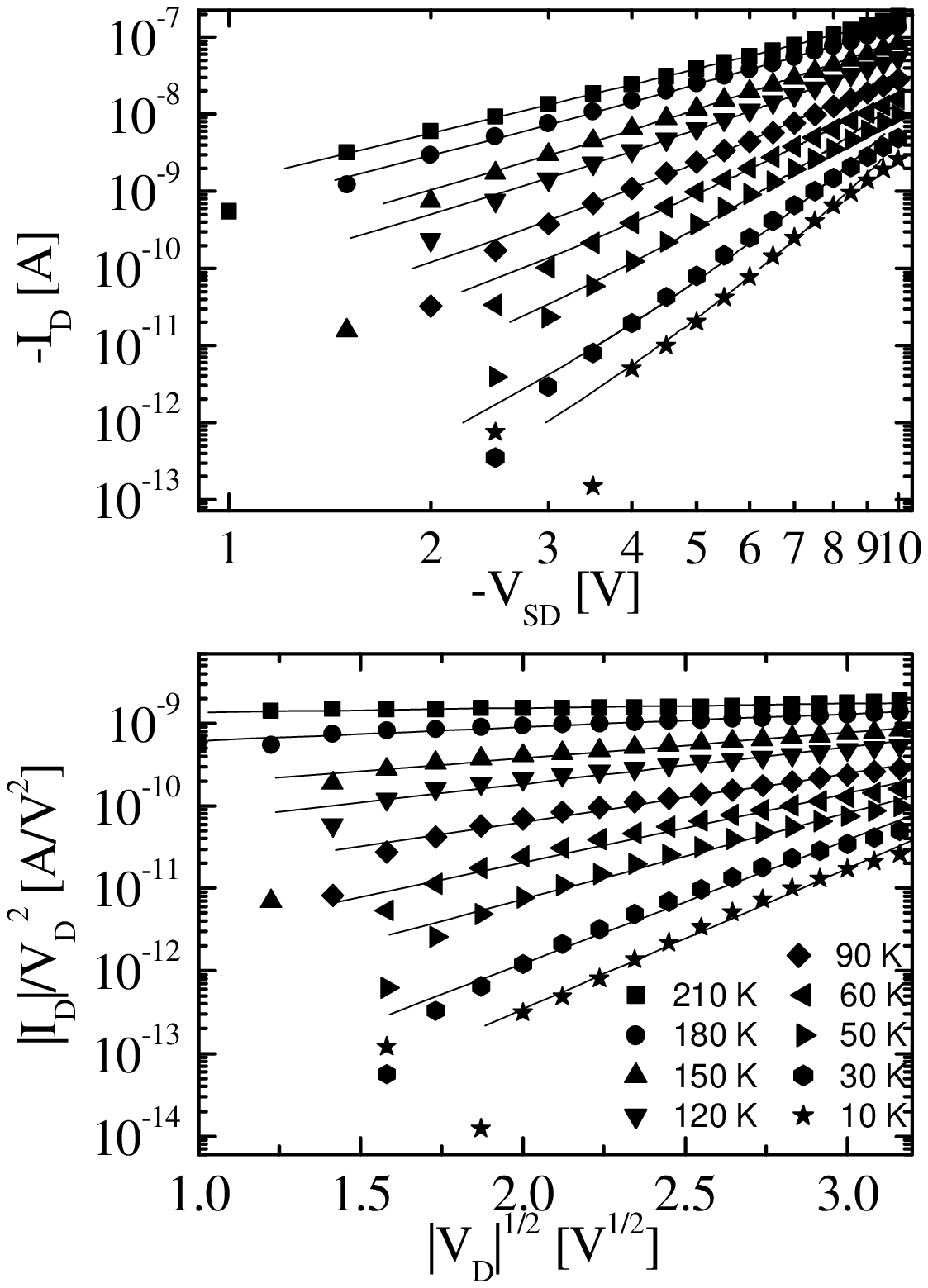}
\end{center}
\vspace{1.5cm}
\caption{of 7, B.H. Hamadani and D. Natelson ``Gated nonlinear transport in organic polymer field effect transistors'', to appear in {\it J. Appl. Phys.}} 
\end{figure} 

\clearpage

\begin{figure} 
\begin{center}
\includegraphics[width=7cm, clip]{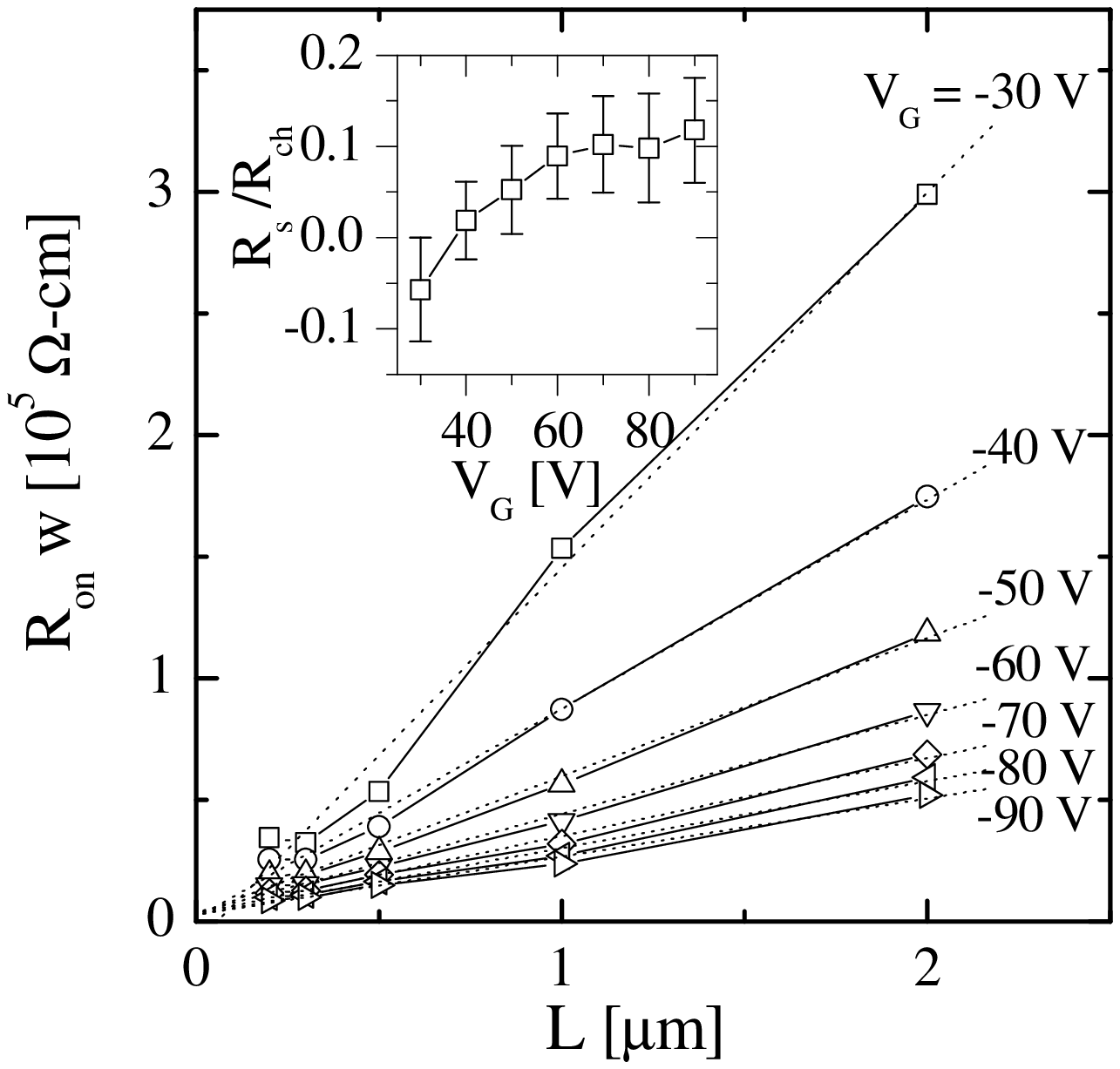}
\end{center}
\vspace{1.5cm}
\caption{of 7, B.H. Hamadani and D. Natelson ``Gated nonlinear transport in organic polymer field effect transistors'', to appear in {\it J. Appl. Phys.}} 
\end{figure} 

\clearpage

\begin{figure} 
\begin{center}
\includegraphics[width=7cm, clip]{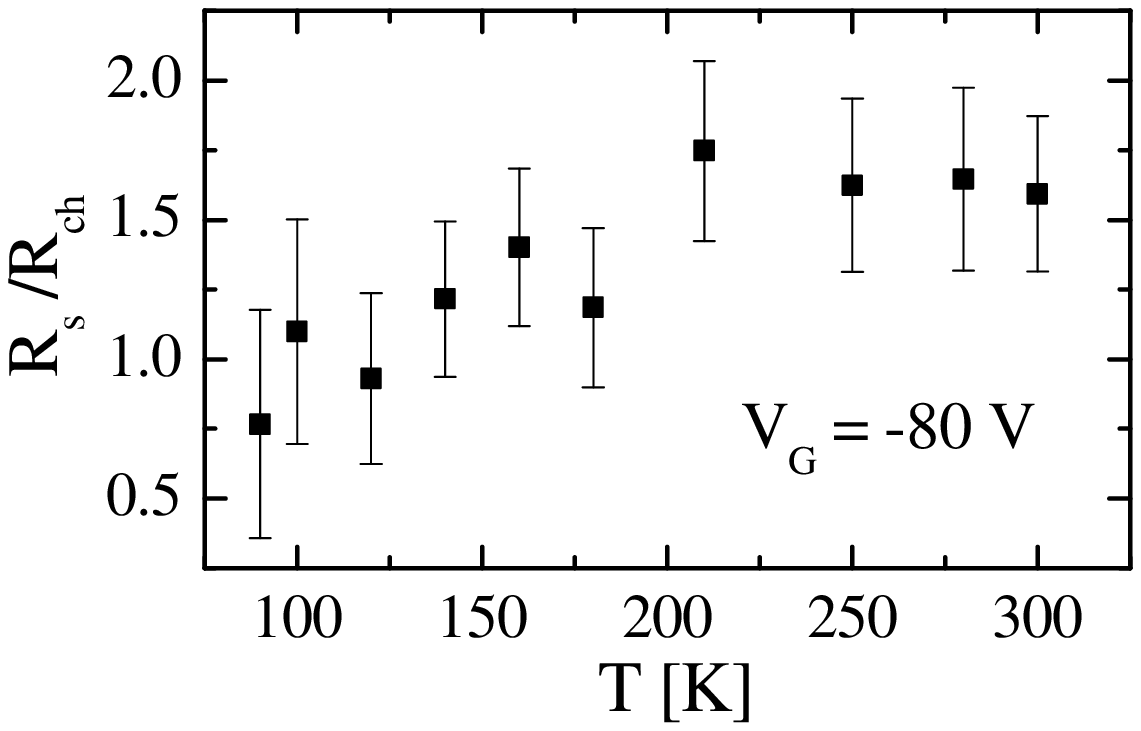}
\end{center}
\vspace{1.5cm}
\caption{of 7, B.H. Hamadani and D. Natelson ``Gated nonlinear transport in organic polymer field effect transistors'', to appear in {\it J. Appl. Phys.}} 
\end{figure} 

\clearpage

\begin{figure} 
\begin{center}
\includegraphics[width=7cm, clip]{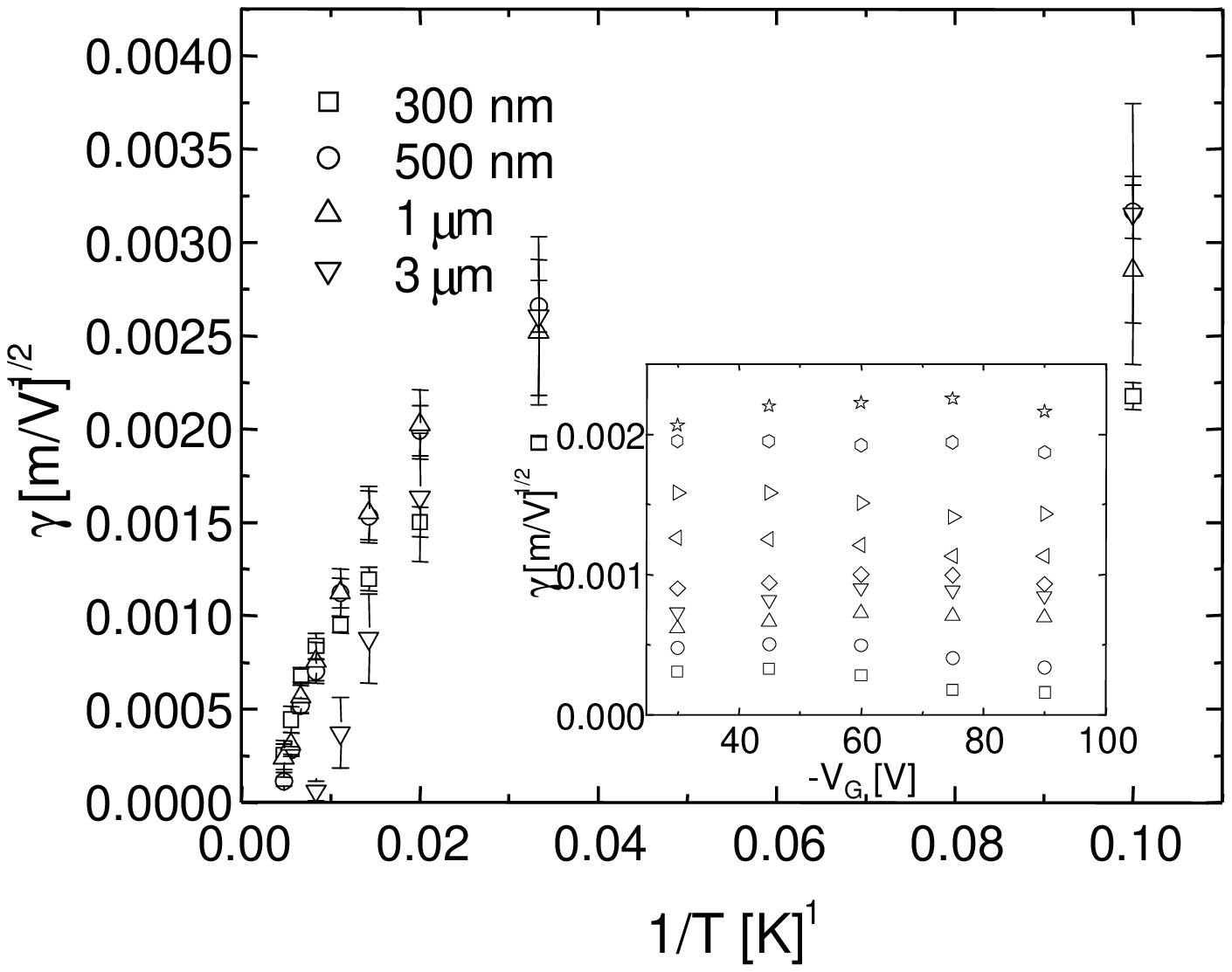}
\end{center}
\vspace{.5cm}
\caption{of 7, B.H. Hamadani and D. Natelson ``Gated nonlinear transport in organic polymer field effect transistors'', to appear in {\it J. Appl. Phys.}} 
\end{figure} 

\clearpage

\begin{figure} 
\begin{center}
\includegraphics[width=7cm, clip]{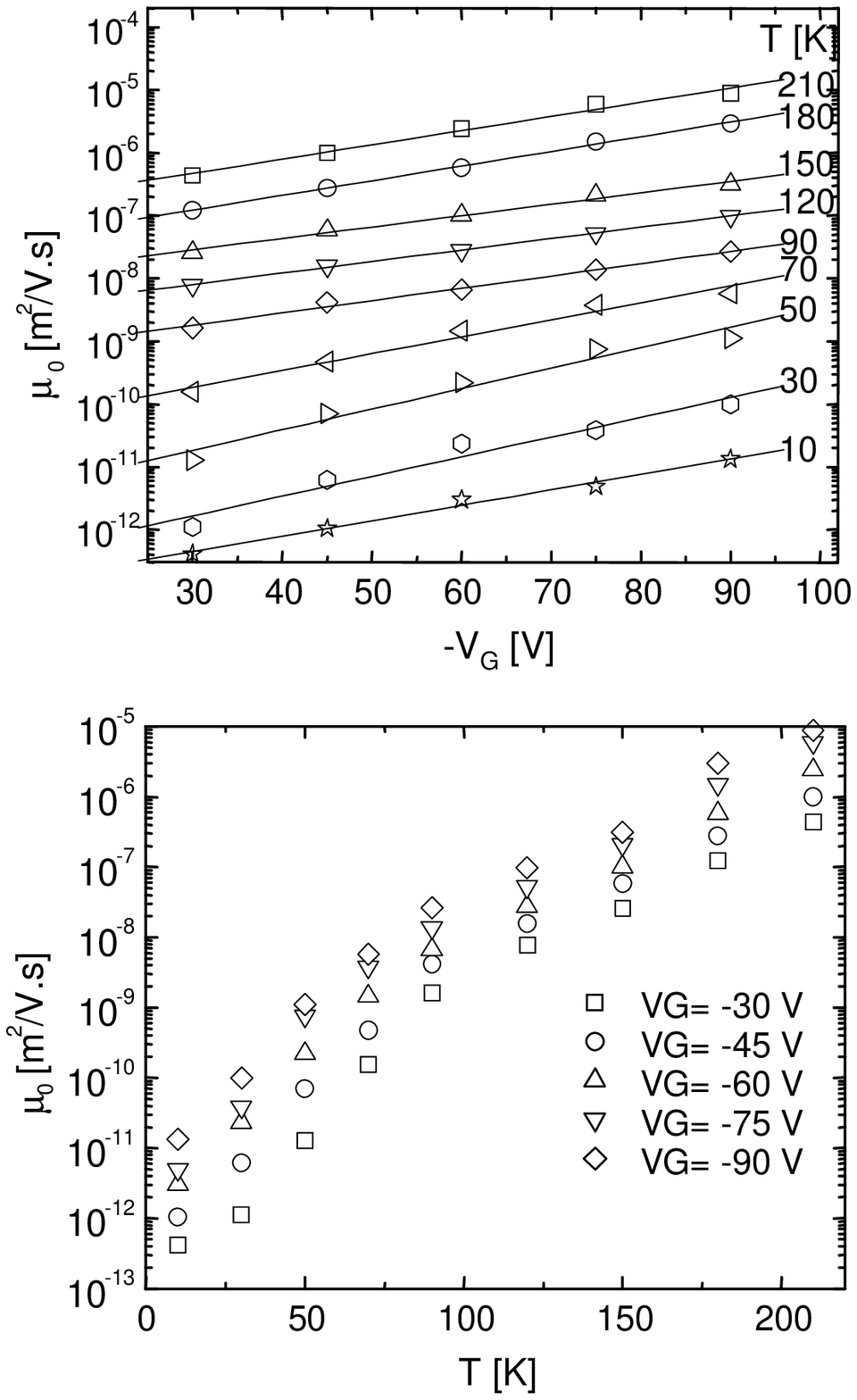}
\end{center}
\vspace{1.5cm}
\caption{of 7, B.H. Hamadani and D. Natelson ``Gated nonlinear transport in organic polymer field effect transistors'', to appear in {\it J. Appl. Phys.}} 
\end{figure}

\end{document}